\documentclass[aps,prd,12pt,preprint,tightenlines,unsortedaddress,
amsfonts,amssymb,amsmath,byrevtex,showpacs,nofootinbib]{revtex4-1}
\usepackage{amssymb,amsmath}
\usepackage{latexsym}
\usepackage{graphicx}
\usepackage{geometry}
\usepackage{epstopdf}
\usepackage{color}
\usepackage{epsfig}

\def\s{\hskip.08em}

\begin{document}
%tightenlines - single space manuscript
%eqsecnum - number equations by section
\newcommand{\dN}{\mathbb N}
\newcommand{\dR}{\mathbb R}
\newcommand{\dC}{\mathbb C}
\newcommand{\dS}{\mathbb S}
\newcommand{\dZ}{\mathbb Z}
\newcommand{\id}{\mathbb I}

\title{Do spikes persist in a quantum treatment \\ of spacetime singularities? }

\author{Ewa Czuchry}
\email{ewa.czuchry@ncbj.gov.pl} \affiliation{Department of Fundamental Research, National Centre for Nuclear Research, Ho{\.z}a 69,
00-681 Warszawa, Poland}

\author{David Garfinkle}
\email{garfinkl@oakland.edu}
\affiliation{Department of Physics, Oakland University, Rochester, Michigan 48309, USA}
\affiliation{Michigan Center for Theoretical Physics, Randall Laboratory of Physics,
University of Michigan, Ann Arbor, Michigan 48109-1120, USA}

\author{John R. Klauder}
\email{john.klauder@gmail.com}
\affiliation{Department of Physics and Department of Mathematics, University of Florida,
Gainesville, Florida 32611-8440,  USA}

\author{W{\l}odzimierz Piechocki}
\email{wlodzimierz.piechocki@ncbj.gov.pl} \affiliation{Department of Fundamental Research, National Centre for Nuclear Research, Ho{\.z}a
69, 00-681 Warszawa, Poland}

\date{\today}

\begin{abstract}
The classical approach to spacetime singularities leads to a simplified dynamics in which spatial derivatives become unimportant
compared to time derivatives, and thus each spatial point essentially becomes uncoupled from its neighbors.  This uncoupled
dynamics leads to sharp features (called ``spikes'') as follows: particular spatial points follow an exceptional dynamical path
that differs from that of their neighbors, with the consequence that, in the neighborhood of these exceptional points, the spatial
profile becomes ever more sharp. Spikes are consequences of the BKL-type oscillatory evolution towards generic singularities of
spacetime. Do spikes persist when the spacetime dynamics is treated using quantum mechanics?  To address this question, we treat
a Hamiltonian system that describes the dynamics of the approach to the singularity and consider how to quantize that system.
We argue that this particular system is best treated using an affine quantization approach (rather than the more familiar methods
of canonical quantization) and we set up the formalism needed for this treatment.   Our investigation, based on
this affine approach, shows the nonexistence of quantum spikes.

\end{abstract}


\maketitle

\section{Introduction}

It is known through the singularity theorems of Penrose, Hawking, and others\cite{HE} that spacetime
singularities are a general feature of gravitational collapse.  However, these theorems give very
little information on the nature of singularities.  It has been conjectured by Belinskii, Khalatnikov,
and Lifshitz (BKL) \cite{BKL,bkl} that, as a spacetime singularity is approached, the dynamics can be well
approximated by neglecting spatial derivatives in the field equations in comparison to time derivatives.
In the course of performing numerical simulations to test the BKL conjecture, Berger and
Moncrief \cite{Gowdyspike} found a strange phenomenon: points at which steep features develop and grow
ever narrower as the singularity is approached.  These features were later named ``spikes''.  Since the
work of \cite{Gowdyspike}, much additional analytical and numerical work has been done on spacetime
singularities (see \cite{BergerReview} for a review), and we now have a good understanding of the
nature of spikes:  (see \cite{dgandbeverly,rendallweaver,dgandweaver,dgsafari,dglimetal,limanduggla,coleyandlim})
rather than being some sort of exception to the BKL conjecture, spikes can be
thought of as a consequence of that conjecture as follows: the neglect of spatial derivatives in the
field equations mandated by the BKL conjecture means that the dynamics at each spatial point is that
of a homogeneous spacetime (albeit a different homogeneous spacetime for each spatial point).
The generic behavior of a homogeneous spacetime consists of a series of epochs, each well
approximated by a different Bianchi I spacetime.  The Bianchi I epochs are connected by short
bounces during which the spacetime is well described by a Bianchi II spacetime.  Though generic
homogeneous spacetimes behave in this way, there are exceptional cases in which the dynamics is
different, remaining in a particular Bianchi I epoch rather than bouncing into the next one.
A spike occurs at a spatial point when the dynamics at that point is of this exceptional sort
while the dynamics of its neighbors is of the generic sort.  The spike point is then stuck in
the old  epoch while all around it, its neighbors are bouncing into the new epoch.

Because spikes depend on exceptional classical dynamics, it is unclear whether they will continue
to exist when the dynamics is treated using quantum theory.  As an analogy, in the upside-down
harmonic oscillator, $x=0$ for all time is a classical solution; but this solution does not
persist in a quantum treatment \cite{Guth,Barton}. Because the BKL conjecture allows the dynamics
of each spatial point to be treated separately, the question of whether spikes persist can
(if the approximation suggested by BKL continues to hold in quantum theory) be treated
just using quantum mechanics rather than quantum field theory or quantum gravity.  Furthermore,
because the exceptional classical dynamics is so delicate as to be easily destroyed, any quantum
destruction of spikes is likely to take place at curvatures much less than the Planck curvature.
Thus, a quantum treatment of spikes is likely to be insensitive to any issues about the ultraviolet
behavior of quantum gravity.

Much of the recent progress on the BKL conjecture comes from treating the Einstein field
equations using a set of scale invariant variables \cite{Uggla,dgprl}.  However, these
treatments are done in terms of field equations rather than Hamiltonian systems, and thus it
is not straightforward to obtain the corresponding quantum dynamics.  To address this difficulty,
Ashtekar, Henderson, and Sloan (AHS) \cite{Ashtekar:2011ck} developed a Hamiltonian system using
variables similar to those in \cite{Uggla,dgprl}.  This new system is designed to address the
BKL conjecture but in a way that one can also perform a quantum treatment.  In this paper, we
will use the system of \cite{Ashtekar:2011ck} to investigate whether spikes persist when treated
using quantum mechanics.

Our paper is organized as follows: In Sec.~II, we solve equations of motion for the natural classical
affine variables and illustrate their temporal behavior leading to classical spikes. In Sec.~III, we
present an alternative quantization process that avoids the need to choose ``Cartesian'' classical
phase space variables to promote to canonical operators and instead supports the quantization of
affine variables with the help of affine coherent states. Section IV~is devoted to the construction
of the physical Hilbert space. Sections V and VI concern the equations of the affine and canonical
quantizations.  In Sec.~VII, we briefly discuss the method of solving the Hamiltonian constraint.
Section VIII presents analytic solutions of the affine constraint equation and concludes that these
solutions do not support the existence of quantum spikes. The last section presents a summary of
our results and indicates how they could be extended using alternative approaches and numerical
methods.

\section{Spikes in the variables of Ashtekar, Henderson, and Sloan}

We begin with a brief description of the variables of Ashtekar, Henderson, and Sloan and refer the reader
to \cite{Ashtekar:2011ck} for the full description.  The approach of \cite{Ashtekar:2011ck} begins with a 
density-weighted triad, its conjugate momentum (which is essentially the extrinsic curvature), and the spatial
connection associated with the triad.  As the singularity is approached, the density-weighted triad is
expected to go to zero, while both the extrinsic curvature and the spatial connection are expected to
blow up. To obtain quantities that are expected to be well behaved at the singularity, AHS define quantities
${{P_i}^j}$ which are contractions of the density-weighted triad with the
extrinsic curvature and ${{C_i}^j}$ which are contractions of the density-weighted triad with the spatial
connection.  In terms of these variables, the BKL conjecture is that as the singularity is approached,
the spatial derivatives of ${{P_i}^j}$ and ${{C_i}^j}$ are negligible compared to their time derivatives;
thus one can consider the dynamics of the ${{P_i}^j}$ and ${{C_i}^j}$ at a single point.  As a consequence
of this form of the BKL conjecture, one finds that the ${{P_i}^j}$ and ${{C_i}^j}$ are symmetric and
can be simultaneously diagonalized; thus the dynamics of these matrices reduces to the dynamics of their
eigenvalues, and \cite{Ashtekar:2011ck} introduces quantities $P_I$ and $C_I$ which are, respectively,
essentially the eigenvalues of ${{P_i}^j}$ and ${{C_i}^j}$.  Thus, for our purposes, the approach to the
singularity is described by a Hamiltonian system consisting of the $C_I$ and $P_I$, as well as any matter
in the spacetime, for which we will use a scalar field $\phi$.  A Hamiltonian system is determined by its
Poisson brackets and its Hamiltonian.  For this system, the Poisson brackets are given by

\begin{equation}\label{a1}
  \{P^I,P^J\}= 0 = \{C_I,C_J\},~~~\{P^I,C_J\}= 2\delta^I_J C_J,~~~\{\phi,\pi\} = 1 \, ,
\end{equation}
while the Hamiltonian (which is also a Hamiltonian constraint) is given \cite{Ashtekar:2011ck} by
\begin{equation}\label{a6}
 H = \frac{1}{2}C^2 - C_I C^I + \frac{1}{2}P^2 - P_I P^I - \frac{\pi^2}{2}= 0\, ,
\end{equation}
which leads to the dynamics
\begin{eqnarray}
% \nonumber to remove numbering (before each equation)
  \dot{P}_I &=& C_I (C-2 C_I), \label{a2} \\
  \dot{C}_I &=& -C_I (P-2 P_I) , \label{a3}\\
  \dot\pi &=& 0, \label{a4} \\
  \dot\phi &=& \pi, \label{a5}
\end{eqnarray}
where $P = P_1 + P_2 + P_3$ and $C= C_1 + C_2 + C_3$.

We now show how spikes form in the vacuum case.  That is, we consider solutions of Eqs. \!(\ref{a6})-(\ref{a5})
with $\pi=0$.
We consider the case with all the $P_I$ positive and order them so that
\begin{equation}
{P_1} > {P_2} > {P_3} \; \; \; .
\label{Porder}
\end{equation}
We assume that at the initial time all the $C_I$ are small enough to be negligible.  Then it
follows from Eq. \!(\ref{a3}),
(\ref{a6}) and (\ref{Porder})
that $C_2$ and $C_3$ are decaying and, therefore, will remain small enough to be negligible.
It then follows from Eqs. \!(\ref{a2}) that $P_2$ and $P_3$
are (to this approximation) constant.  Thus, we only need to find the time development of
$C_1$ and $P_1$.  With $C_2$ and $C_3$ negligible, Eqs. \!(\ref{a2}) and (\ref{a6}) become
\begin{eqnarray}
{{\dot P}_1} = - {{({C_1})}^2}\, ,
\label{P1dot}
\\
- {{({C_1})}^2} = 2 ({P_1 ^2} + {P_2 ^2} + {P_3 ^2}) - {P^2}
\label{ham2}\, .
\end{eqnarray}
However, Eq. \!(\ref{ham2}) can be written as
\begin{equation}
- {{({C_1})}^2} = ({P_1} - {P_+})({P_1} - {P_-})\, ,
\label{quad}
\end{equation}
where the constants $P_\pm$ are given by
\begin{equation}
{P_\pm} = {P_2} + {P_3} \pm 2 {\sqrt {{P_2}{P_3}}}\, .
\end{equation}
We, therefore, find that Eq. \!(\ref{P1dot}) becomes
\begin{equation}
{{\dot P}_1} =  ({P_1} - {P_+})({P_1} - {P_-})\, .
\label{P1dot2}
\end{equation}
Let $P_{10}$ be the value of $P_1$ at the initial time $t_0$.  Then it follows from Eq. \!(\ref{P1dot2})
that
\begin{equation}
{\frac {{P_+}-{P_1}} {{P_1}-{P_-}}} = {\frac {{P_+}-{P_{10}}} {{P_{10}}-{P_-}}} \; {\exp [({P_+}-{P_-})(t-{t_0})]}\, .
\label{P1soln}
\end{equation}
Let $C_{10}$ be the value of $C_1$ at time $t_0$.  Then it follows from Eq. \!(\ref{quad}) that
\begin{equation}
{P_{10}} = {\textstyle {\frac 1 2}} \left ( {P_+}+{P_-} + {\sqrt {{{({P_+}-{P_-})}^2} - 4 {C_{10} ^2}}} \right )\, .
\end{equation}
Now define the function $f(t)$ by
\begin{equation}
f(t) = {\frac {2{C_{10}}} {{P_+}-{P_-} + {\sqrt {{{({P_+}-{P_-})}^2} - 4 {C_{10} ^2}}}}} \;
{\exp [{\textstyle {\frac 1 2}}({P_+}-{P_-})(t-{t_0})]}\, .
\label{fdef}
\end{equation}
Then it follows from Eqs. \!(\ref{P1soln})--(\ref{fdef}) using straightforward algebra that
\begin{equation}
{P_1} = {\frac {{P_+} + {P_-}{f^2}} {1 + {f^2}}}\;\; .
\label{P1soln2}
\end{equation}
It then follows from Eq. \!(\ref{quad}) that
\begin{equation}
{C_1} = ({P_+}-{P_-}) {\frac f {1+{f^2}}}\;\; .
\label{C1soln}
\end{equation}

Now consider the case where ${C_{10}} \ne 0$.  By the assumption that $C_1$ is small at the
initial time, it follows that at
that time $f \ll 1$.  However, from the exponential factor in Eq. \!(\ref{fdef}) it follows
that for sufficiently
late times we have $f \gg 1$.
It then follows from Eq. \!(\ref{P1soln2}) that initially ${P_1} \approx {P_+}$ but at late
times $P_1 \approx {P_-}$.  That is, there is a bounce where
$P_1$ goes from $P_+$ to $P_-$.  It also follows from Eq. (\ref{C1soln}) that $C_1$ is small
at both early and late times and is only non-negligible during the bounce.

\begin{figure}
\centering
\includegraphics[width=0.5\textwidth]{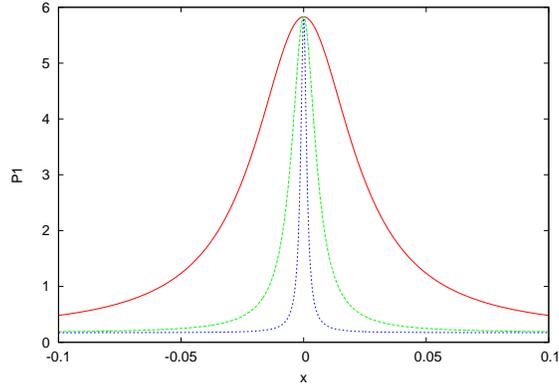}
\caption{$P_1$  vs $x$ at $t=3$ (red), $t=3.5$ (green), and $t=4$ (blue)}
\label{fig:P1}
\end{figure}
\begin{figure}
\centering
\includegraphics[width=0.5\textwidth]{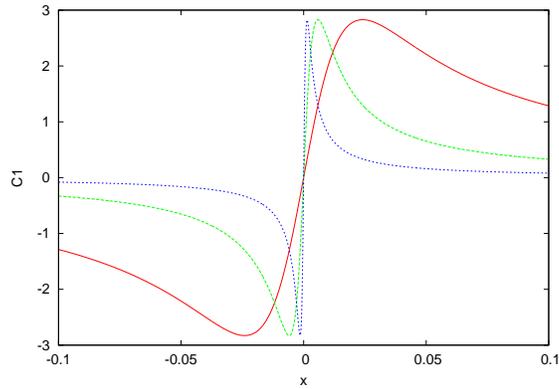}
\caption{$C_1$  vs $x$ at $t=3$ (red), $t=3.5$ (green), and $t=4$ (blue) }
\label{fig:C1}
\end{figure}

Now consider the case where ${C_{10}}=0$.  Then $f=0$ for all times and, thus, it follows that
${P_1}={P_+}$ and ${C_1}=0$.

Finally, consider the dependence on spatial position.  Suppose that at the initial time there
is a region where $C_1$ is positive and another region where $C_1$ is negative.  Define a spatial
coordinate $x$ such that $x=0$ is the boundary between the two regions.  Then by continuity, we
have that ${C_{10}}=0$ at $x=0$.  Therefore, we find that for all times ${P_1}={P_+}$ at $x=0$
while for all other points eventually we have ${P_1}={P_-}$.  The closer to $x=0$ a spatial
point is, the smaller the value of $C_{10}$ and, therefore, the longer a time it takes until $f$ at
that point becomes large.  Thus, though eventually all points near $x=0$ bounce, it takes longer
for the nearer points to bounce.  Thus, at a given time, a graph of $P_1$ vs $x$ will show a
peak at $x=0$ and that curve will become more and more steep as time goes on.  This is the spike.

As an illustration, consider the case ${C_{10}} = \epsilon x $ with $\epsilon = 0.05$ and take
${P_2}=2, \, {P_3}=1$ and
${t_0} =0$.  Figure 1 shows $P_1$ vs $x$ at $t=3, \, t=3.5$ and $t=4$, while Fig. 2
shows $C_1$  vs $x$ at those same times.

\section{Enhanced Quantization}

Besides canonical quantization, which is discussed in Sec.~IV, we begin with a very different quantization procedure that avoids the
problematic procedure of choosing the right set of canonical variables to promote to canonical operators. The nature of the classical
problem features variables that lead to an affine Lie algebra, which is then incorporated in the quantum formulation as affine coherent
states. This quantization method is also useful for addressing the issue of the existence of spikes at a semiclassical level.

\subsection{Affine algebra}

Initially, we propose to quantize the classical system \eqref{a1}--\eqref{a6}
by making use of the affine
coherent states quantization method (see \cite{JRK} and references therein).
We begin with some remarks about use of the affine variables in the classical formulation of the problem. To connect with
notation that is more common for affine variables,
we make the partial redefinition $(C_I, P^J)=: (C_I, -2D^J) $,
which turns the system \eqref{a1}--\eqref{a6}
into the traditional Poisson bracket affine formulation,
\begin{equation}\label{b1}
  \{D^I,D^J\}= 0 = \{C_I,C_J\},~~~\{C_J, D^I\}= \delta^I_J C_J \, ,
\end{equation}
which is called the affine Lie algebra. For the scalar field, we adopt conventional canonical coordinates with the standard Poisson bracket
\begin{equation}\label{b2}
  \{\phi,\pi\} = 1 \, .
\end{equation}
For this problem the classical Hamiltonian is constrained to be zero  \cite{Ashtekar:2011ck}, and it is given by
\begin{equation}\label{b7}
H = \frac{1}{2}C^2 -  \sum_I C_I^2 + 2 D^2 -  4 \sum_I D_I^2 - \frac{1}{2}\pi^2 = 0\, ,
\end{equation}
where $D = D_1 + D_2 + D_3$ and $C= C_1 + C_2 + C_3$.
Thus, the dynamics takes the form
\begin{eqnarray}
% \nonumber to remove numbering (before each equation)
  \dot{D}_I &=& C_I (C_I - \frac{1}{2}C), \label{b3} \\
  \dot{C}_I &=& 2C_I (D - 2 D_I) , \label{b4} \\
  \dot\pi &=& 0, \label{b5} \\
  \dot\phi &=& \pi . \label{b6}
\end{eqnarray}

Unlike the traditional momentum, which serves to translate the canonical coordinate $C_I$, the variable $D^I$ serves to dilate $C_I$. Thus the affine algebra
divides into three sectors: (1) $C_I < 0$, (2) $C_I > 0$, and (3) $C_I=0$. The first two types are quite similar, while the third type is relatively trivial.
Consequently, we will concentrate on types (1) and (2). Thus, it is convenient to define the  principal sectors in the  kinematical phase space as
\begin{eqnarray}
 \label{b8} \Pi_-^I &:= &\{(C_I, D^I) \;|\; C_I \in \dR_- , D^I \in \dR \}, \\
 \label{b8b} \Pi_+^I &:=& \{(C_I, D^I) \;|\; C_I \in \dR_+ , D^I \in \dR \}.
\end{eqnarray}

\subsection{ Kinematical Hilbert space}

There are two principal, inequivalent, irreducible self-adjoint representations of the Lie algebra \eqref{b1}
corresponding to the  sectors \eqref{b8} and  \eqref{b8b}.
They are defined by the affine quantization principle: $C_I \rightarrow \hat{C}_I$
and $D^I \rightarrow \hat{D}^I$, such that
\begin{equation}\label{b88}
[\hat{C}_I, \hat{C}_J] = 0 = [\hat{D}^I, \hat{D}^J],~~~~[\hat{C}_J, \hat{D}^I] = i\hbar \;\delta^I_J \hat{C}_J \; .
\end{equation}
The operators $\hat{C}_I$ and $\hat{D}^I$ are conveniently represented by
\begin{eqnarray}\label{c1}
\hat{D}^I f (x^I) &:= &-i \hbar/2 \;(x^I\;\partial/\partial x^I +  \partial/\partial  x^I\; x^I)\;f(x^I)\nonumber\\
  & =&-i \hbar \;(x^I\;\partial/\partial x^I + 1/2)\;f(x^I),~~~\\
\hat{C}_I f(x^I)&:=& x^I f(x^I) \label{c11}\, ,
\end{eqnarray}
where $f\in \mathcal{H}^I := span \{L^2(\dR_-,dx^I) \oplus L^2(\dR_+,dx^I)\}$, and where $I = 1, 2, 3$.

For quantization of the scalar field algebra \eqref{b2}
 we use the canonical variables and the following representation
\begin{equation}\label{c2}
\hat{\pi} g (\phi) := -i \hbar \;\frac{\partial}{\partial \phi} g(\phi),~~~ \hat{\phi}g(\phi):=  \phi g(\phi)\, ,
\end{equation}
where $g \in \mathcal{H}_\phi :=L^2 (\dR, d\phi)$, so that $[\hat{\phi},\hat{\pi}] = i \hbar\; \id$.

The {{\it kinematical}  %\it physical}
Hilbert space $\mathcal{H}$ of the entire system is defined to be
\begin{equation}\label{c3}
\mathcal{H}:= span \{\mathcal{H}^1\otimes \mathcal{H}^2\otimes  \mathcal{H}^3\otimes \mathcal{H}_\phi\} \, ,
\end{equation}
which takes into account the usual quantum entanglement of all degrees of freedom.

\subsection{Construction of affine coherent states}

It is important to observe that the classical Hamiltonian treats the three $C$ variables, as well as the three
$D$ variables, in identical fashion in that the Hamiltonian is invariant if the several variables are
permuted among themselves. This feature of symmetry is worth preserving in introducing the coherent
states for these variables.
Thus, the irreducible components of the {\it affine}
coherent states corresponding to each of the two sectors $\Pi_- =\bigcup_I \Pi_- ^I$ and $\Pi_+
=\bigcup_I \Pi_+ ^I$, are defined
as follows

\begin{eqnarray}
% \nonumber to remove numbering (before each equation)
\label{c4} |p,q,-\rangle &:=& \prod_I |p^I,q_I,I, -\rangle\nonumber\\
&:=& e^{i \sum_I p^I\; \hat{C}_I /\hbar} e^{-i \sum_I \ln (|q_I|/\mu)\; \hat{D}_I /\hbar} |\eta, - \rangle
~~~\mbox{for sector}~~~ \Pi_- \, ,\\
\label{c5} |p,q,+\rangle &:=& \prod_I |p^I,q_I,I, +\rangle \nonumber\\
& :=&e^{i \sum_I p^I\; \hat{C}_I /\hbar} e^{-i \sum_I \ln (q_I/\mu)\; \hat{D}_I /\hbar}|\eta, + \rangle
~~~\mbox{for sector}~~~ \Pi_+ \, ,
%\label{c6} |p,q,0\rangle &=& ... |\eta, 0 \rangle~~~\mbox{for sector}~~~ \Pi_0^I \, ,
\end{eqnarray}
where $p:=(p^1, p^2,p^3)$ and $q:=(q_1, q_2, q_3)$. The   so-called  fiducial vectors $|\eta,
- \rangle=\otimes_I |\eta, I - \rangle$ and $|\eta, +
\rangle=\otimes_I |\eta, I + \rangle$
are defined by the equations
\begin{eqnarray}
% \nonumber to remove numbering (before each equation)
\label{c7} [(\hat{C}_I/\mu) + 1 - i \hat{D}_I/ (\hbar\beta)] |\eta, I - \rangle&=& 0 \; ,\\
\label{c8} [(\hat{C}_I/ \mu) - 1 + i \hat{D}_I/ (\hbar\beta)] |\eta, I + \rangle&=& 0 \; ,
% \label{c9} [...] |\eta, 0 \rangle&=& 0 \; ,
\end{eqnarray}
where $\mu>0$ and $\beta>0$ denote two free parameters chosen the same for each set of variables $\hat{C}_I$ and
$\hat{D}_I$. (It is also useful to regard $\tilde{\beta}\,[\equiv \hbar\beta]$ and $\hbar$ as two separate
parameters for each $I$, especially for approaching the classical limit.) The role of $\mu$ and $\beta$ can
be seen in the expressions

\begin{eqnarray}
\label{cc7} \langle x| \eta, I +\rangle&=& M x^{-1/2} (x/\mu)^{\beta} e^{-\beta (x/\mu)},~~~~ 0 <x< \infty\; ,\\
\label{cc8} \langle x| \eta, I -\rangle&=& M |x|^{-1/2} (|x|/\mu)^{\beta} e^{-\beta (|x|/\mu)},~~~~ -\infty< x< 0\; ,
\end{eqnarray}
where $M$ denotes a factor to secure normalization, e.g., $\langle \eta, I \pm|\eta, I \pm\rangle=1$.
It follows that
\begin{equation}\label{d1}
\langle \eta, \pm
| \hat{C}_I | \eta, \pm
\rangle = \pm
\mu\;,~~~\langle \eta, \pm
| \hat{D}_I | \eta, \pm
\rangle = 0 \; .
\end{equation}

It may happen that the appropriate affine coherent states for our study involve a direct sum of the $+$ and $-$
irreducible versions, such as

\begin{equation}
|p,q, I\rangle :=\theta (q) |p,q, I, +\rangle \oplus \theta (-q)|p,q, I, -\rangle \;,
\end{equation}
where $\theta (y):=1$ if $y>0$ and $\theta (y):= 0$ if $y<0$.  In order to incorporate both the positive and
negative spectrum cases for $\{\hat{C_I}\}$, we shall use the direct sum of vectors, $|\eta\rangle :=
|\eta, +\rangle \oplus |\eta, -\rangle$, in what follows.
% In such a case, it follows that
%$\langle \eta | \hat{C}_I |\eta \rangle =0$ for all $I$.

In addition to the affine coherent states, we introduce {\it canonical} coherent states for the scalar field,
which are defined by
\begin{equation}\label{d2}
|\pi, \phi\rangle := e^{-i \phi \hat{\pi}/\hbar} e^{i \pi \hat{\phi}/\hbar} |\alpha\rangle \; ,
\end{equation}
where the fiducial vector $|\alpha\rangle$ is chosen (modulo a phase factor) to be the solution to the equation
\begin{equation}\label{d3}
(\omega \hat{\phi} + i \hat{\pi}) |\alpha\rangle = 0\;,
\end{equation}
in which $\omega$ is a free positive parameter. It follows that
\begin{equation}\label{d4}
\langle \pi,\phi| \hat{\pi} |\pi, \phi\rangle = \pi\;,~~~\langle \pi,\phi| \hat{\phi} |\pi, \phi\rangle= \phi \; .
\end{equation}

We choose states $|y\rangle$ (previously called $|\phi\rangle$ and which are renamed here to avoid conflicting
notation), where $\langle y|y'\rangle =\delta(y-y')$ and $-\infty<y, y'<\infty$, so that $\langle y|\,\hat{\phi}
=y\,\langle y|$ as well as $\langle y|\,\hat{\pi}=-i\hbar \partial/\partial y \langle y|$. Thus,

\begin{equation}
(\omega y+\hbar\partial/\partial y) \langle y|\alpha\rangle = 0
\end{equation}
leads eventually to (with the usual meaning of $\pi$ in the first factor)
\begin{equation}
\langle y|\pi,\phi\rangle= (\omega/\pi \hbar)^{1/4}\,\exp(i\pi y/\hbar-\omega (y-\phi)^2/2\hbar)\;.
\end{equation}
This last equation asserts that $\langle y|\pi,\phi\rangle$ is the $y$-representation of the coherent state;
likewise, it follows that $\langle \pi,\phi| y\rangle\,[\,=\langle y|\pi,\phi\rangle^*\,]$ is the coherent-state
representation of the $y$-state.

Introducing
\begin{equation}\label{d5}
W_\pm:=\prod_{I=1}^3 |\, \langle \eta,\pm| ~\hat{C}_I^{- 1} | \eta, \pm \rangle |\; ,
\end{equation}
we can get a resolution
of unity for each of the two sectors $\Pi_-$ and $\Pi_+$:

\begin{equation}\label{d6}
\int |\pi, \phi\rangle \langle \pi,\phi|\otimes|p,q,\pm\rangle \langle p,q,\pm |
\frac{d\pi\;d\phi\;d^3p\;d^3q}{ h^4 W_\pm } = \id_\pm \;
\end{equation}
provided that $W_\pm < \infty$ (where $h$ denotes Planck's constant).

\subsection{Enhanced classical action functional}

The use of coherent states as part of the classical/quantum connection is related to the restricted variation
of vectors in the quantum action functional only to appropriate coherent states, which then leads to the
enhanced classical action functional, for which $\hbar>0$ throughout. We next spell out this connection.

The quantum action functional is given by
\begin{equation}
A_Q= \int_0^T \langle\psi(t)| [ i\hbar(\partial/\partial t)-\hat{H} ] |\psi(t)\rangle\, dt
\end{equation}
and leads to the Schr\"odinger equation when general variations are admitted. However, if the variations are
limited to coherent states---including just the variations that a macroscopic observer could make---it follows
\cite{JRK} that the so-restricted quantum action functional becomes (with summation on $I$ implied)

\begin{eqnarray}
A_{Q(R)}&=&\int_0^T\langle p(t),q(t); \pi(t),\phi(t)| [i\hbar(\partial/\partial t)- \hat{H} ] |p(t),q(t);
\pi(t),\phi(t)\rangle\,dt\nonumber\\
&=& \int_0^T \{\, -q_I(t)\,\dot{p}^I(t) +\pi(t)\, \dot{\phi}(t) - H(p(t),\pi(t);q(t),\phi(t))\,\}\label{abc5} \; dt \;,
\end{eqnarray}
where $|p,q;\pi,\phi\rangle := (|p,q, + \rangle \otimes |\pi,\phi\rangle )\oplus (|p,q, - \rangle \otimes |\pi,\phi\rangle) $,
which according to the principles of {\it enhanced quantization} \cite{JRK} can be viewed as the
enhanced classical action functional in which $\hbar$ retains its normal positive value. The relation of
the quantum Hamiltonian to the expression

\begin{equation}\label{abc6}
H(p,\pi;q,\phi):=\langle p,q;\pi,\phi|\,\hat{H}(\hat{C}_I,\hat{D}_I; \hat{\pi},\hat{\phi})|p,q;\pi,\phi\rangle
\end{equation}
is known as the {\it Weak Correspondence Principle} \cite{k15}.
We can use this relationship to help choose the quantum Hamiltonian $\hat{H}$.

An affine quantization that includes the Weak Correspondence Principle does {\it not} involve the assumption that
the classical coordinates must be ``Cartesian coordinates'' as is the case for canonical quantization. This is because
in enhanced quantization the variables $p$ and $q$ are {\it not} ``promoted'' to operators in the quantization process.
This difference ensures that enhanced quantization can provide different physics than that offered by
canonical quantization.

It follows that (no summation intended)
\begin{eqnarray}
H(p,\pi;q,\phi)&=& \langle p,q; \pi,\phi|\,\hat{H}(\hat{C}_I, \hat{D}_I; \hat{\pi}, \hat{\phi})\,|p,q;\pi,\phi \rangle \\
&=& \langle \eta; \alpha|\,\hat{H}( (q_I/\mu)\,\hat{C}_I, \hat{D}_I + p^I (q_I/\mu) \hat{C}_I;\hat{\pi}+\pi, \nonumber
\hat{\phi}+\phi)\,|\eta; \alpha\rangle\;,
\end{eqnarray}
where $|\eta; \alpha\rangle := |\eta\rangle \otimes |\alpha\rangle$.
If we adopt the naive form of the quantum operator $\hat{H}$ suggested by conventional canonical
quantization---applied to the classical Hamiltonian \eqref{b7}
---the result leads to [with $\langle (\cdots)\rangle:=\langle\eta; \alpha| (\cdots) |\eta; \alpha\rangle$
in what follows]

\begin{eqnarray}
H(p,\pi;q,\phi)&=&\frac{1}{2}\langle [\sum_I q_I (\hat{C_I}/\mu)]^2 \rangle -\sum_I q_I^2 \langle (\hat{C}_I/\mu)^2\rangle\nonumber\\
& & ~~~\, +2 \langle (\sum_I[ \hat{D}_I+p_I(q_I/\mu) \hat{C}_I])^2\rangle \nonumber \\
& & ~~~\, -4 \sum_I \langle (\hat{D}_I+p_I(q_I\hat{C}_I/\mu)^2\rangle -\frac{1}{2} \langle[ \hat{\pi}+\pi^2 ] \rangle \label{dde}\;,
\end{eqnarray}
which may be written in the form

\begin{eqnarray}
H(p,\pi;q,\phi)&=& \frac{1}{2}(\sum_I q_I)^2 - \sum_{I} q_I^2 + 2 \sum_{I,J} p_I q_I p_J q_J -4 \sum_I p_I^2q_I^2 -\frac{1}{2}\pi^2 \nonumber\\
& & ~~~\, +\frac{1}{2} \sum_{I.J} q_Iq_J [ \langle(\hat{C_I}/\mu)(\hat{C_J}/\mu) \rangle-1]-\sum_I q_I^2 [\langle (\hat{C}_I/\mu)^2\rangle-1]\nonumber\\
& & ~~~\,+ 2 \sum_{I,J} p_I q_I p_J q_J\,[\langle[(\hat{C}_I/\mu)(\hat{C}_J/\mu)]\rangle-1]
-4 \sum_I p_I^2q_I^2[\langle ( \hat{C}_L/\mu)^2 \rangle-1]\nonumber\\
& & ~~~ \, + 2 \langle (\sum_I \hat{D_I})^2\rangle- 4\sum_I \langle \hat{D_I}^2\rangle-\frac{1}{2} \langle\hat{\pi}^2\rangle
\label{ddd1}\;.
\end{eqnarray}
We note that the variables $q_I$ and $p_I$ are related to the former classical variables  according
to the relations: $q_I:=C_I$ and $p_Iq_I:=D_I$.
Thus, the first line in \eqref{ddd1}
is the classical Hamiltonian \eqref{b7},
while all the terms in the three remaining lines in \eqref{ddd1} are
$O(\hbar)$ (based on using the parameters $\tilde{\beta}\,[:=\hbar\beta]$ and $\hbar$). These terms constitute
quantum corrections to the classical Hamiltonian generated by the enhanced quantization point of view.

The last line in \eqref{ddd1} are constants and can be canceled by subtracting them from $\hat{H}$. The terms
on lines two and three involve quantum corrections to line one and are dealt with by adopting the enhanced
Hamiltonian given by

\begin{eqnarray}
H(p,\pi;q,\phi)&=&\frac{1}{2}\langle [\sum_I q_I (\hat{C_I}/\mu)]^2 \rangle -\sum_I q_I^2 \langle (\hat{C}_I/\mu)^2\rangle \label{ddg}\\
& & +2 \langle [\sum_I p_I q_I (\hat{C}_I/\mu)]^2\rangle
-4 \sum_I p_I^2q_I^2 \langle(\hat{C}_I/\mu)^2\rangle -\frac{1}{2} \pi^2 \nonumber  \;,
\end{eqnarray}

The terms $\langle (\hat{C}_I/\mu)\rangle=\pm1$ while $\langle (\hat{C}_I/\mu)^2\rangle=1+z$, where

\begin{eqnarray}
&&z:=-1+\nonumber\\
&&\;\int_{-\infty}^\infty |(x/\mu)|^{2\tilde{\beta}/\hbar+1}\,e^{-(2\tilde{\beta}/\hbar)|(x/\mu)|}\;dx\Big{/}
\int_{-\infty}^\infty |(x/\mu)|^{2\tilde{\beta}/\hbar-1}\,e^{-(2\tilde{\beta}/\hbar)|(x/\mu)|}\;dx\;,
\nonumber\\ \end{eqnarray}
which shows that $z=O(\hbar)$. Equation \eqref{ddg} ensures that the enhanced classical Hamiltonian is very
much like the traditional classical Hamiltonian, and its enhanced classical equations of motion involve small corrections to the
traditional classical equations of motion.

\section{Passing to the physical Hilbert space}

The constraint of the Hamiltonian vanishing is an essential requirement in the quantum theory as it was in
the classical theory. This has the effect of reducing the {\it kinematical} Hilbert space to the {\it physical}
Hilbert space. In other words, we propose to follow the Dirac quantization scheme \cite{PAM,HT}:  first, quantize
(in the  kinematical Hilbert space)  then  second, introduce the constraints (to identify the  physical Hilbert space).
We realize this scheme with the help of reproducing kernel Hilbert spaces (see, e.g., \cite{Klauder:2006rs, NA} ).

\subsection{Reproducing kernel Hilbert space}

The essence of reproducing kernel Hilbert spaces is readily explained. For example, as we have seen, the
kinematical Hilbert space is spanned by the set of coherent states $|p,q;\pi,\phi\rangle$. Thus, every vector
in that space is given by
\begin{equation}\label{rk1}
|\Psi\rangle = \sum_k a_k \;|p_k,q_k;\pi_k,\phi_k\rangle \;,
\end{equation}
provided that
\begin{equation}\label{rk2}
0\le \langle \Psi|\Psi\rangle=\sum_{j,k} a^*_ja_k \langle p_j,q_j;\pi_j,\phi_j|p_k,q_k;\pi_k,\phi_k\rangle <\infty \;.
\end{equation}
Observe that the set of coherent states, $\{|p,q;\pi,\phi\rangle\}$, forms a continuously labeled set of vectors,
which spans the kinematical Hilbert space, but whose elements are, therefore,  not linearly independent as in a conventional basis set.
Instead, the set of coherent states represents a kind of ``continuous basis'' for a separable Hilbert space.

Next, we give a functional representation for every abstract vector by introducing

\begin{equation}\label{rk3}
\Psi(p,q;\pi,\phi):=\langle p,q;\pi,\phi|\Psi\rangle =\sum_k a_k\;\langle p,q;\pi,\phi|p_k,q_k;\pi_k,\phi_k\rangle\;.
\end{equation}
Another vector is given by its functional representation as follows
\begin{equation}\label{rk4}
\Phi(p,q;\pi,\phi)=\langle p,q;\pi,\phi|\Phi\rangle=\sum_{j'} b_{j'}\;\langle p,q;\pi,\phi|p_{j'},q_{j'};\pi_{j'},\phi_{j'}\rangle\;,
\end{equation}
where the set $\{p_{j'},q_{j'};\pi_{j'},\phi_{j'}\}$ for $|\Phi\rangle$ is generally different from the set $\{p_k,q_k;\pi_k,\phi_k\}$ for $|\Psi\rangle$.
In the reproducing kernel Hilbert space, the inner product of two such functional representation elements is given by

\begin{equation}\label{rk5}
(\Phi,\,\Psi) := \sum_{j',k} b^*_{j'}a_k \langle p_{j'},q_{j'};\pi_{j'},\phi_{j'}|p_k,q_k;\pi_k,\phi_k\rangle\;,
\end{equation}
which is just a functional representative of $( \Phi,\Psi)=\langle \Phi|\Psi\rangle$.

Observe that the inner product of two coherent states, $\langle p_{j'},q_{j'};\pi_{j'},\phi_{j'}|p_k,q_k;\pi_k,\phi_k\rangle$,
 serves as a reproducing kernel; if the vector
$\langle \Phi|$ is chosen as the vector $\langle p,q;\pi,\phi|$ (i.e., $b_{j'}=\delta_{j',1})$, then the result
of the inner product ``reproduces'' the expression for $\langle p,q;\pi,\phi|\Psi\rangle$. Traditionally, the reproducing
kernel is chosen as jointly continuous in both arguments. In our case, the
reproducing kernel using coherent states is automatically jointly continuous
because the group properties of the affine and canonical groups ensure continuity. Hence, like all reproducing kernel
Hilbert spaces, our reproducing kernel Hilbert space is composed entirely of  continuous functions.

\subsection{Coherent state overlap as a reproducing kernel}

The foregoing discussion is based on the general theory of reproducing kernel Hilbert spaces. However, when suitable coherent states
generate the reproducing kernel, as in the present case, some additional properties hold true. In particular, there is an equivalent, second procedure
for the inner product of two functional representatives. Equation \eqref{d6} shows that a suitable integral over projection operators
onto coherent states leads to the unit operator in the kinematical Hilbert space. Choosing the positive section as an example, the general coherent state matrix elements of that equation lead to the equation
\begin{equation} \label{dd1}
   \langle p'',q'';\pi'',\phi''|p',q';\pi',\phi'\rangle = \int \langle p'',q'';\pi'',\phi''|p,q;\pi,\phi\rangle
 \langle p,q;\pi,\phi|p',q';\pi',\phi'\rangle\;d\mu(p,q;\pi,\phi) \;,
 \end{equation}
 where $d\mu$ represents the integration measure in \eqref{d6}. It follows from this equation that the inner product of the two functional representatives
 $\Phi(p,q;\pi,\phi)=\langle p,q;\pi,\phi|\Phi\rangle$ and $\Psi(p,q;\pi,\phi)=\langle p,q;\pi,\phi|\Psi\rangle$ is given by
\begin{equation} \label{dd2}
   \langle\Phi|\Psi\rangle=\int \Phi^*(p,q;\pi,\phi)\,\Psi(p,q;\pi,\phi)\;d\mu(p,q;\pi,\phi)\;.
 \end{equation}
 In particular, if $\langle\Phi|=\langle p'',q'';\pi'',\phi''|$, this relation leads to an example of the reproducing kernel property. Indeed, if one lets $f(p,q,\pi,\phi)$ be a general element of the space $L^2(\mathbb{R}^8)$, the reproducing kernel acts as a projection operator onto a valid vector in the
 kinematical Hilbert space, e,g.,
 \begin{equation} \label{dd3}
  \Psi_f(p',q';\pi',\phi')=\int\langle p',q';\pi',\phi'|p,q;\pi,\phi\rangle\, f(p,q,\pi,\phi)\;d\mu(p,q;\pi,\phi)\;.
  \end{equation}
  It may well be that dealing with this integral version of the inner product is more appropriate in special cases.

\subsection{Projection operators for reducing the kinematical Hilbert space}
\def\E{{\rm I}\hskip-.2em{\rm E}}

Let $\E$ represent a {\it projection operator} (hence, $\E^2=\E^\dagger =\E$). If $\E$ is smaller then the
unit operator, it follows that $\langle p,q;\pi,\phi|\,\E\,|p',q';\pi',\phi'\rangle$ serves as a reproducing
kernel for a {\it subspace} of the original Hilbert space. In particular, we suppose that $\E$ is a projection
operator onto the subspace where the Hamiltonian vanishes, i.e., $\E=\E(\hat{H}=0)$. The Hamiltonian operator
consists of two parts one with $\hat{C}_I$ and $\hat{D}_I$ and the other with $\hat{\pi}^2/2$. Let us assume
that the first part of $\hat{H}$ has a discrete spectrum $\{E_n\ge0\}$ and that the second part has a
continuous spectrum $0\le y^2/2<\infty$. Thus, the eigenfunctions $|E_n;y\rangle = |E_n\rangle \otimes |y\rangle$,
satisfy
$\hat{H}\,|E_n; y\rangle =(E_n-y^2/2)\,|E_n; y\rangle$ and $\langle E_n ;x | E_m  ;y\rangle = \delta_{nm}\delta (x-y)$.
Suppose  the full spectrum of $\hat{H}$ implies that  $\Sigma_n\int_{0}^\infty |E_n;y\rangle\langle E_n;y| \;dy$
is the unit operator. In such a case we have

\begin{equation}\label{abc}
\langle p,q;\pi,\phi|\E|p',q';\pi',\phi'\rangle=\sum_n \langle p,q|E_n\rangle\langle E_n|p',q'\rangle \,
\langle \pi,\phi|\sqrt{2E_n}\rangle \langle \sqrt{2E_n}|\pi',\phi'\rangle\;.
\end{equation}
It follows that Eq. \eqref{abc} defines   a valid representation of a reproducing kernel that includes only
the subspace where $\hat{H}=0$. Therefore, a functional representation for every vector $\Psi_{\E}$ of our physical Hilbert is given by
\begin{equation}\label{aBc}
  \Psi_{\E} (p,q;\pi,\phi) = \langle p,q;\pi,\phi|\Psi_{\E}\rangle = \sum_k a_k\; \langle p,q;\pi,\phi|\E|p_k,q_k;\pi_k,\phi_k\rangle \; .
\end{equation}

\def\F{{\rm I}\hskip-.2em{\rm F}}

Operators for the kinematical Hilbert space lead to generally different operators for the physical Hilbert space. Since the affine
coherent state vectors $\{|p,q;\pi,\phi\rangle\}$ span the kinematical Hilbert space, it follows that the projected coherent state
vectors $\{ \E |p,q;\pi,\phi\rangle \}$ span the physical Hilbert space, as described above. In like fashion, an operator $\hat{A}$
that applies to the kinematical Hilbert space leads to an operator $\hat{A}_{\E} := \E \hat{A} \E$ that applies to the physical Hilbert space.
Sometimes a general  property of $\hat{A}$ is not preserved by $\hat{A}_{\E}$, such as being self adjoint. If $\hat{A}\ge0$, then
$\hat{A}_{\E}\ge0$ as well, and if $\hat{A}$ is self adjoint then $\hat{A}_{\E}$ can also be chosen to be self adjoint. On the other hand,
if $\hat{Q}$ and $\hat{P}$ (with $[\hat{Q},\hat{P}]=i\hbar \,\id$) are both self adjoint and a projection operator $\F$ is such that
$\F \hat{Q} \F$ is self adjoint and strictly positive,
then it follows that $\F \hat{P} \F$ can never be self adjoint. This is just the situation that is overcome  by choosing the affine
variables $\hat{Q}$ and $\hat{D}:=(1/2)[\hat{Q}\hat{P}+\hat{P}\hat{Q}]$ (with $[\hat{Q},\hat{D}]=i\hbar \hat{Q}$) for which $\hat{Q}>0$
and $\hat{D}$ are both self adjoint.

Note that elements of the physical Hilbert space enjoy the same integral representation of inner products as noted earlier since
\begin{equation} \label{dd4}
   \langle\Phi|\E|\Psi\rangle=\int \Phi_{\E}^*(p,q;\pi,\phi)\,\Psi_{\E}(p,q;\pi,\phi)\;d\mu(p,q;\pi,\phi)\;,
 \end{equation}
 where, as before, $\Psi_{\E}(p,q;\pi,\phi):=\langle p,q;\pi,\phi|\E| \Psi\rangle$.

 The development with time $(t)$ in the kinematical Hilbert space follows traditional expressions, such as if $\hat{H}(t)$ denotes the
(possibly time dependent) Hamiltonian operator that acts on an operator $\hat{A}(t)$ for which the time dependence is only that caused by
the Hamiltonian, i.e., for which $\partial A(t)/\partial t = 0$, then the Heisenberg equation of motion $i\hbar\, d\hat{A}(t)/dt= [\hat{A}(t),\hat{H}(t)]$ holds as usual. On the other hand, for the physical Hilbert space, one must impose the projection operator
{\it after} forming the commutator such as $i\hbar\, \E d\hat{A}(t)/dt \E=\E[\hat{A}(t),\hat{H}(t)]\E$ and not by imposing the projection
operator {\it before} forming the commutator in the form $i\hbar\, d\E \hat{A}(t)\E/dt= [\E \hat{A}(t)\E,\E \hat{H}(t)\E]$. Not only does
the latter equation involve a different number of projection operators $(\E)$
on each side of the equation, but, as we expect in the current problem, the physical Hilbert space is such that $\E \hat{H}(t)\E=0$.
Consequently, for the former equation of motion, the operators $\E \hat{A}(t)\E$ evolve properly within the physical Hilbert space for
suitable choices of the operator $\hat{A}(t)$.

It is noteworthy that the energy eigenstates for the first part of the Hamiltonian (i.e., only with
$\hat{C}$ and $\hat{D}$) are degenerate leading
to the possibility that there may be various energy eigenstates for a single energy value. This is likely
to be true as well for the energy value $E=0$. Thus, there could be a family of zero-energy eigenstates
for which
$\hat{\pi}^2/2$ is not required to ensure that $\hat{H}\,|m: (E=0) \rangle = 0$, $m\in\{1,2,3,...\}$.
In such a case only affine
coherent states, $|p,q\rangle$, are necessary and no canonical coherent states, $|\pi,\phi\rangle$, are
needed. To find the states $|m: (E=0)\rangle$
requires solving the zero-energy Schr\"odinger equation $\hat{H}|m: (E=0)\rangle = 0$.  It is important to understand
that the form of the differential equation leading to zero-energy solutions in the canonical quantization scheme in
the following section is entirely different from the differential equation leading to zero-energy solutions in the affine quantization
scheme as the latter equation is shown in the following section. Besides that difference in formulation, there is one
advantage that an affine quantization offers in that the proper subtraction terms can be decided so that the enhanced
classical Hamiltonian has the form given in \eqref{ddg} such that, even when $\hbar>0$, the enhanced classical solutions
follow when the enhanced classical Hamiltonian is constrained to vanish.

\section{Affine Quantization}

Let us try to define $\hat{H}$ by making use of the classical form of $H$ defined by
Eq. \!\eqref{b7}.
Since there are no products of $C_I$ and $D_I$ in  \eqref{b7}, and due to \eqref{b88}, the mapping of the Hamiltonian
 \eqref{b7} into a Hamiltonian operator  is straightforward.  We get
  \begin{eqnarray}
% \nonumber to remove numbering (before each equation)
 \hat{H}  &=&\frac{1}{2}(\sum_I x_I)^2        %%%H = \frac{1}{2}C^2 -  \sum_I C_I^2 + 2 D^2 -  4 \sum_I D_I^2 - \frac{1}{2}\pi^2 = 0\, ,
 -\sum_I x_I^2 - 2\hbar^2 [\sum_I\,(x_I\frac{\partial}{\partial x_I}+1/2)\,]^2   \nonumber \\
 & &\hskip3em + 4\hbar^2\sum_I\,(x_I\frac{\partial}{\partial x_I}+1/2)^2 +\frac{1}{2}\hbar^2\frac{\partial^2}{\partial\phi^2}  \nonumber \\
  &=& \hbar^2\big(-\frac{3}{2}
+2 \sum_I x_I^2\frac{\partial^2}{\partial x_I^2} - 4 \sum_{I<J} x_I x_J \frac{\partial^2}{\partial x_I
\partial x_J}  \big)
 \nonumber \\
& &\hskip3em  +  \sum_{I<J} x_I x_J - \frac{1}{2} \sum_I x_I^2 + \frac{\hbar^2}{2} \frac{\partial^2}{\partial\phi^2} =: \hat{H}_g + \hat{H}_\phi\; ,
\label{r2}\end{eqnarray}
where $\hat{H}_\phi = \frac{\hbar^2}{2} \frac{\partial^2}{\partial\phi^2},$  and where $\hat{H}_g$ is the gravitational contribution.

One can show (see Appendix ) that the operator $\hat{H}$ is Hermitian  on a dense subspace of $L^2 (\dR^3, d^3 x)$ of the functions
satisfying suitable boundary conditions.

\section{Canonical quantization}

Though we think that the form of the Poisson brackets given in Eqs. \!(\ref{a1}) indicates that our system is best treated with affine quantization methods, we nonetheless briefly consider how this system might be treated using the more usual canonical quantization methods. Recall that in canonical quantization one begins with classical configuration variables $X_I$ and momentum variables $P_I$ having Poisson brackets,
\begin{equation}\label{inv}
\{P^I,X_J\}= {\delta^I_{\,J}}
\end{equation}
One then realizes the kinematical Hilbert space as $L^2 (\dR,dX_I)$ and the operator $P^I$ as
${P^I}= i \hbar\, (\partial /\partial {X_I})$.

Now consider the case where all $C_I$ are positive and define the $X_I$ by ${X_I}=(1/2) \ln {C_I}$.
Then it follows from Eq. \!(\ref{a1}) that $P^I$ and $X_I$ satisfy the canonical Poisson bracket given in
Eq. \!(\ref{inv}).

The Hamiltonian constraint [Eq. \!(\ref{a6})] written in terms of $X_I$ then becomes
\begin{eqnarray}
{{\left ( {e^{2{X_1}}} + {e^{2{X_2}}} + {e^{2{X_3}}} \right ) }^2} - 2 \left ( {e^{4{X_1}}} + {e^{4{X_2}}} +
{e^{4{X_3}}} \right )
\nonumber
\\
+ {{\left ({P_1} + {P_2} + {P_3}\right ) }^2} - 2 \left ( {P_1 ^2} + {P_2 ^2} + {P_3 ^2}
\right )  - {\pi ^2} = 0 \, .
\label{hamX}
\end{eqnarray}

The physical Hilbert space is obtained by replacing
$P_I$ by $i \hbar\, (\partial /\partial {X_I})$ and then imposing the Hamiltonian constraint as an operator acting on the wave function $\psi$. 
We, thus, obtain the following equation:
\begin{equation}\label{a7}
{\hbar ^2} {\frac {{\partial ^2} \psi} {\partial {\phi ^2}}} =  {\hbar ^2} \left (\sum_{I\neq J} \frac {\partial ^2 \psi}
{\partial {X_I} \partial {X_J}} - \sum_I \frac{\partial^2 \psi}{\partial X_I^2} \right )
+ \left( \sum_I {e^{4{X_I}}} - \sum_{I\neq J} e^{2({X_I}+{X_J})}\right) \psi \; .
\end{equation}

The rhs of Eq. \!\eqref{a7} defines an Hermitian operator on a dense subspace of $L^2(\dR^3, d^3 X)$ of the functions
satisfying suitable boundary conditions.

\section{Methods of imposing the Hamiltonian constraint}

In order to find the quantum fate of spikes, we will need to impose the Hamiltonian constraint, possibly using numerical methods, and examine the properties of the resulting wave function $\psi$.  Note that in ordinary quantum mechanics the Hamiltonian operator generally involves the Laplacian, and the energy eigenvalue equation (``time-independent Schr\"odinger equation'') is an elliptic equation.  However, it is a general property of quantum cosmology that the quantum Hamiltonian constraint equation is a hyperbolic equation.  (This strange property is essentially due to the conformal degree of freedom of the metric behaving differently from the other metric degrees of freedom.)  In contrast to elliptic equations, which lead to boundary value problems, hyperbolic equations lead to initial value problems.  To pose the initial value problem, one must choose a timelike coordinate $T$ and choose initial data on a surface of constant $T$.

For the case of canonical quantization and the imposition of  \eqref{a7}, a convenient choice of coordinates is the following:

%Written in this form, it is not clear what sort of equation this is.  However, we now introduce a change of variables that makes clear that eqn. %(\ref{a7}) is hyperbolic.  Define the variables $T, Y$ and $Z$ by
\begin{eqnarray}
T := {X_1} + {X_2} + {X_3}\label{a8} \, ,
\\
Y := {X_1} - {X_2}\label{a9} \, ,
\\
Z := {\textstyle {\frac 1 2}} ({X_1} + {X_2}) - {X_3}\label{a10} \, ,
\end{eqnarray}
which turns \eqref{a7} into
%Then eqn. (\ref{a7}) becomes
\begin{eqnarray}
&{\hbar ^2}& \left ( - {\frac {{\partial ^2} \psi} {\partial {T^2}}} + {\textstyle {\frac 4 3}}
{\frac {{\partial ^2} \psi} {\partial {Y^2}}} + {\frac {{\partial ^2} \psi} {\partial {Z^2}}} +
{\textstyle {\frac 1 3}}{\frac {{\partial ^2} \psi} {\partial {\phi^2}}}\right )
\nonumber
\\
&+&  \frac{1}{3}\,{e^{(4/3)(T+Z)}} \left [ 2 \left ( 1 + {e^{Y-2Z}} + {e^{-(Y+2Z)}} \right )
- \left ( {e^{2Y}} + {e^{-2Y}} + {e^{-4Z}} \right ) \right ]\psi = 0 \; ,
\label{a11}
\end{eqnarray}
where $\psi \in L^2(R^4, dT dY dZ d\phi)$.
Equation \eqref{a11} has explicitly a hyperbolic form, suitable for numerical simulations,  with $T$ playing the role of an evolution parameter.

For the case of affine quantization, the Hamiltonian defined by Eq. \!\eqref{r2} yields an equation analogous to Eq. \!\eqref{a7}, which
is defined in  $L^2(R^4, d^3x d\phi)$ and reads
\begin{equation}\label{r3}
{\hbar ^2} {\frac {{\partial ^2} \psi} {\partial {\phi ^2}}} =  4 {\hbar ^2} \left(\sum_{I\neq J} x_I x_J \frac {\partial ^2 \psi}
{\partial {x_I} \partial {x_J}} - \sum_I x^2_I\frac{\partial^2 \psi}{\partial x_I^2} + \frac{3}{4}\psi \right)
+ \left( \sum_I x^2_I - \sum_{I\neq J} x_I x_J\right) \psi \; .
\end{equation}
The solution $\psi$ of Eq. \!\eqref{r3} has, potentially, a very different physical interpretation than that of the solution of Eq. \!\eqref{a7}.

One can diagonalize Eq. \eqref{r3} in a similar way as  Eq.  \!\eqref{a7}.
Introducing the  variables,

\begin{align}
T:=& x_1 x_2 x_3,\\
Y:=& \frac{x_1}{x_2},\\
Z:=& \frac{\sqrt{x_1 x_2}}{ x_3} ,
\end{align}
enables rewriting  \eqref{r3} in the following form:
\begin{align}
&  4 {\hbar ^2} \left( -{T^2\frac {{\partial ^2}\psi } {\partial {T^2}}} +{\textstyle {\frac 4 3}}
{Y^2\frac {{\partial ^2}\psi } {\partial {Y^2}}} + {Z^2\frac {{\partial ^2}\psi } {\partial {Z^2}}} + \frac{1}{12}{\frac {{\partial ^2}\psi } {\partial {\phi^2}}} - 4T\frac{\partial\psi}{\partial T} +{\textstyle {\frac 4 3}}Y\frac{\partial\psi}{\partial Y}+Z\frac{\partial\psi}{\partial Z} - \frac{3}{4}\psi\right)\nonumber\\
&+ {\frac 1 3}\,{(TZ)^{2/3}} \left [ \left ( {Y^{2}} + {Y^{-2}} + {Z^{-4}} \right )-2 \left ( 1 + \frac{Y}{Z^2}+\frac{1}{YZ^2} \right ) \right ]
 \psi = 0\; .\label{r4}
\end{align}
In this hyperboliclike equation, suitable for numerical simulations, the variable $T$ plays the role of an evolution parameter.

\section{Exploring the Affine Constraint Equation}

We recall the affine Hamiltonian constraint equation \eqref{r3} and set the left-hand side to zero seeking a solution
with only $\{x_I\}$ variables. It follows that a ``near
solution'' to the resulting constraint equation is given by
\begin{equation}\label{jrk1}
    \Psi(x_1,x_2,x_3):=(2\hbar)^{-3/2}\,\exp\{-(1/2\hbar)[\sum_I\,|x_I|\s]\s\} \;,
\end{equation}
and, as presented,  $\Psi$ is normalized, i.e., $\int |\Psi(x_1,x_2,x_3)|^2 \,d^3\!x=1$. If we now put the remaining
zero-point energy [appearing as $3\hbar^2$
in \eqref{r3}] as part of the
original Hamiltonian, this solution satisfies the equation $\hat{H}\,\Psi=0$, and, thus,  \eqref{jrk1} represents a solution
of the quantum constraint. At first sight it seems strange that a function that has a discontinuous derivative---thanks to
$|x_1|$, etc.---can satisfy the modified \eqref{r3}. In fact, all  eight independent solutions of the modified affine
Hamiltonian constraint \eqref{r3} have a similar form given by
\begin{equation}\label{jrk2}
   \Psi(x_1,x_2,x_3; J_\pm):= (\hbar)^{-3/2}\,\Pi_I\, \{\,[J_{+,I}\theta(x_I)+J_{-,I}\theta(-x_I)]\, e^{-(1/2\hbar)\,|x_I| }\,\}\;,
\end{equation}
where $\theta(y):=1$ for $y>0$ and $\theta(y):=0$ for $y<0$, and $|J_{+,I}|^2+|J_{-,I}|^2=1$ for each $I$. This form of the
wave function contains finite jumps
at $x_I=0$ when $|J_{+,I}|\not=|J_{-,I}|$, for one or more $I$. The solution \eqref{jrk2} is valid even though there are terms
of the form $x_I^2\,\delta\,'(x_I)$ as well as
$x_Ix_J\,\delta(x_I)\,\delta(x_j)$ for $I\not=J$, all of which vanish. The factor $8\,[=2^3]$ arises from the variety available
from the eight inequivalent terms $J_{\pm,1}\,J_{\pm,2}\,J_{\pm,3}$. Hereafter, to simplify the notation in this section, we
assume that the plain symbol $\Psi$ (or $\Phi$) denotes any vector in the eight-dimensional physical Hilbert space with the form
given in \eqref{jrk2}.

It is noteworthy that certain operators can be simplified when they are confined to act on vectors in the physical Hilbert
space. Clearly, the relation $\hat{C}_I\,\Psi=x_I\,\Psi$ holds, and it follows that
\begin{equation}\label{jrk3}
\hat{D}_I\,\Psi=-i\hbar[x_I\,\partial/\partial x_I+1/2] \,\Psi=(i/2)[\,x_I-\hbar]\,\Psi=(i/2)[\,\hat{C}_I-\hbar]\,\Psi\;,
\end{equation}
which shows that the action of $\hat{D}_I$ is effectively multiplicative in nature. Indeed, it follows that
\begin{equation}
  \hat{D}_I\, \hat{C}^p_I \,\Psi=\{\, [ \hat{D}_I,\, \hat{C}^p_I] + \hat{C}^p_I\,\hat{D}_I\}\,\Psi=\{- i \hbar \, p\,\hat{C}^p_I+(i/2)\hat{C}^p_I\,[\hat{C}_I-\hbar]\}\,\Psi\;.
  \end{equation}

  Although these equations are correct, however, it follows that while $\Psi$ is a vector in the physical Hilbert space, it
  is a fact that $\hat{C}^p_I\,\Psi$,
   for $p>0$,  is  not a vector in the physical Hilbert space. To address that situation, we can obtain the part of that
   vector in the physical Hilbert space by taking the inner product with another vector $\Phi$ in the physical Hilbert space,
   which leads to $(\Phi, \hat{C}^p_I\,\Psi)$. In the
   eight-dimensional physical Hilbert space, this inner product, for $I=1$, becomes
   \begin{eqnarray}
    &&(\Phi, \hat{C}^p_1\,\Psi) =\hbar^{-1}\,\int\, [K_{+,1}^*\theta(x_1)+K_{-,1}^*\theta(-x_1)]\nonumber\\
    &&\hskip8em \times\,[J_{+,1}\theta(x_1)+J_{-,1}\theta(-x_1)]
   \, \,x^p_1\,e^{-|x_1|/\hbar}\,dx_1 \nonumber\\
   &&\hskip4.5em =p!\,\hbar^{p}\,[K_{+,1}^*J_{+,1}+(-1)^p K_{-,1}^*J_{-,1}]\;,
    \end{eqnarray}
    where $J_{\pm,1}$ refers to $\Psi$ and $K_{\pm,1}$ refers to $\Phi$.

This simplification of the form taken by the operator $\hat{D}_I$ leads to a simplification of the equation of motion.
The classical equations
$\dot{C}_I=C_I[D_I-2D]$ transform, for the kinematical Hilbert space, to the operator equation
\begin{equation}\label{jrk4}
\dot{\hat{C}}_I= (1/2)\hat{C}{_I}\,[\hat{D}_I-2\hat{D}]+(1/2)[\hat{D}_I-2\hat{D}]\,\hat{C}{_I}\;.
\end{equation}
To fit it into the physical Hilbert space, this equation becomes
\begin{equation}\label{jrk5}
(\Phi,\,\dot{\hat{C}}_I\,\Psi)= (1/2)\,(\Phi,\,\{\,\hat{C}{_I}\,[\hat{D}_I-2\hat{D}]+[\hat{D}_I-2\hat{D}]\,\hat{C}{_I}\}\,\Psi)\;,
\end{equation}
which becomes an equation involving only the $\{\hat{C}_I\}$ operators, namely
\begin{equation}
(\Phi,\,\dot{\hat{C}}_I\,\Psi)= (1/2)\,(\Phi,\,\{\, -i\hbar\,\hat{C}_I +i\hat{C}_I(\hat{C}_I-\hbar)+2\sum_L\,[\,i\hbar\,\hat{C}_L-i\hat{C}_L\,
(\hat{C}_L-\hbar)]\,\} \,\Psi)\, .
\end{equation}

It follows that higher-order time derivatives of $\hat{C}$ can be developed leading to an expression of the form
\begin{equation}
   \hat{C}_I(t) =   \hat{C}_I(0) +t\,\dot{\hat{C}}_I(0)+(t^2/2)\,\ddot{\hat{C}}_I(0)+...\;,
   \end{equation}
   which leads to  the general expression given by
   \begin{equation}
   (\Phi,\,\hat{C}_I(t)\,\Psi)=(\Phi,\, M_I(t, \hat{C}_L(0))\,\Psi)
   \end{equation}
   which introduces the time dependent, $8\times8$ matrix, $M_I(t,\hat{C}_L(0))$.

   Observe that the matrix $M_I$ itself does not depend on any specific vector in the physical Hilbert space. The vectors
   in the physical Hilbert space
   are distinguished by the factors $J_{\pm,I}$ that signify the coefficients that define a given vector $\Psi$. To discuss
   position dependence in real space, as was the case in Sec.~II in order to study the position dependence of potential
   spikes, we let $\bar{x}_1$ represent a position in real space. For us, dealing with the physical Hilbert space, the position
   parameter appears in the choice of the parameters $J_{\pm,I}$ in the physical Hilbert space vectors; the position parameter
   $\bar{x}_1$ does {\it not} appear in the matrix $M_I(t,\hat{C}_L(0))$.
    Let us first choose the initial position of a vector to be at position $\bar{x}_1=0$ in space. This we can accommodate by
   choosing $J_{+,1} =J_{-,1}=1/\sqrt{2}$. If, instead, we want to be at a small, nonzero position $0<\bar{x}_1<\hbar$, we can choose $J_{+,1}=\sqrt{(1+\bar{x}_1/\hbar)/2}$ and $J_{-,1}=\sqrt{(1-\bar{x}_1/\hbar)/2}$ so that
   \begin{equation}
   (\hbar)^{-1}\int [|J_{+,1}|^2\theta(x_1)+|J_{-,1}|^2\theta(-x_1)]\,x_1\,\exp[-|x_1|/\hbar\,]\,dx_1=\bar{x}_1\;.
    \end{equation}

  Using the several tools discussed above, let us consider some general properties of the expression for $(\Psi(\bar{x}_1),\,\hat{C}(t)\,\Psi(\bar{x}_1)\,)$, focusing on the   influence of the space position $\bar{x}_1$. With $\Psi(0):=\Psi(\bar{x}_1=0)$, we observe that
   \begin{eqnarray}
    &&\hskip-2em |\; (\Psi(\bar{x}_1),\,\hat{C}_1(t)\,\Psi(\bar{x}_1)\,) - (\Psi(0),\,\hat{C}_1(t)\,\Psi(0)\,)\,|\nonumber \\
     &&=|\; (\Psi(\bar{x}_1),\,\hat{C}_1(t)\,\Psi(\bar{x}_1)\,) - (\Psi(0),\,\hat{C}_1(t)\,\Psi(0)\,)\nonumber\\
       &&\hskip3em  +(\Psi(0),\,\hat{C}_1(t)\,\Psi(\bar{x}_1)\,) - (\Psi(0),\,\hat{C}_1(t)\,\Psi(\bar{x}_1)\,)\,| \nonumber\\
    &&\le |(\Psi(\bar{x}_1),\,\hat{C}_1(t)\,\Psi(\bar{x}_1)\,) - (\Psi(0),\,\hat{C}_1(t)\,\Psi(\bar{x}_1)\,)\,|\nonumber\\
      &&\hskip3em  +|(\Psi(0),\,\hat{C}_1(t)\,\Psi(\bar{x}_1)\,) - (\Psi(0),\,\hat{C}_1(t)\,\Psi(0)\,)\,|\\
      &&\le ||\Psi(\bar{x}_1)-\Psi(0)||\,||\hat{C}_1(t)\,\Psi(\bar{x}_1||+||\Psi(\bar{x}_1)-\Psi(0)||\,||\hat{C}_1(t)^\dagger\,
      \Psi(0)||\;,\nonumber
       \end{eqnarray}
where $||\Psi||$ denotes the norm of the vector $\Psi$. Finally, with the vectors $\Psi$ normalized to unity, we find that
  \begin{eqnarray}
   |\; (\Psi(\bar{x}_1),\,\hat{C}_1(t)\,\Psi(\bar{x}_1)\,) - (\Psi(0),\,\hat{C}_1(t)\,\Psi(0)\,)\,|
    \le 2\,||\Psi(\bar{x}_1)-\Psi(0)||\,\,||\hat{C}_1(t)|| \,,\nonumber \\
   \end{eqnarray}
 in which $||\hat{C}_1(t)||$ now denotes the operator norm of the $8\times8$ matrix representation of the physical Hilbert space
 form, i.e., $M_1(t,\hat{C}_L(0))$,  of the given operator. Moreover, this equation provides a bound on the hypothetical quantum
 spike, and with the temporal and spacial portions bounded and completely separated, we believe that quantum spikes do not exist.
 In other words, these solutions do not support the existence of quantum spikes since they prohibit the temporal and spatial behavior characteristic of the classical spike behavior.

To achieve the eight solutions of the quantum Hamiltonian constraint we had to subtract a numerical term that was proportional to $\hbar^2$. This is not unlike using normal ordering to find solutions of a quantum problem, e.g., removing the zero-point energy in a free field when it is composed of a set of harmonic oscillators. Although our problem has been treated as a quantum mechanics problem, it should be appreciated that such a problem applies to every spatial point and, thus, the overall zero-point energy diverges. The solutions we have obtained for the physical Hilbert space are its least energy states simply because they do not cross the axis and change sign just as ground-state wave functions traditionally behave.

\section{Conclusions}

We have set up a formalism to treat the question of whether spikes persist in a quantum treatment of spacetime singularities.
We argue that a promising method for addressing this question is to treat the quantum dynamics of individual spatial points
using the Hamiltonian system of Ashtekar, Henderson, and Sloan.  We further note that the form of the Poisson brackets of this
system indicates that the affine approach to quantization would be more natural for this system than the usual canonical
quantization method.  As shown in the previous section, an exploration of the physical Hilbert space using the
affine analysis leads to the conclusion that quantum spikes do not exist.

We now consider particular ways to apply the formalism developed in this paper to extend our results on the effect of
quantum mechanics on spikes.  Recall from Sec. II that classical spikes occur because the dynamics at a particular point
(the center of the spike) is of an exceptional sort, different from the dynamics of all neighboring points.  Thus, the question
of whether quantum effects destroy spikes is essentially the question of whether quantum effects destroy these exceptional
classical trajectories in the physical phase space.  Though quantum corrections are small (at least far from the Planck
scale) nonetheless, the unstable nature of the exceptional trajectories means that they might be destroyed by even such small
effects.  The simplest form of this question is to retain the classical phase space, but to replace the classical Hamiltonian
with the enhanced Hamiltonian of Sec. III D, and to see whether this change alone is enough to destroy the exceptional
trajectory. More generally, we would consider wave packets that start out peaked around the exceptional classical trajectory
and see whether quantum uncertainty makes those wave packets spread so that at later times they are no longer peaked around
the classical trajectory.  These wave packets would need to satisfy the Hamiltonian constraint that the wave function is
annihilated by the quantum Hamiltonian operator (Eq. (\ref{r4}) in the affine case or (\ref{a11}) for the canonical case).
Since wave packets are known to have tendency to spread during an evolution, we propose to examine this issue quite
independently by making use of the reproducing kernel Hilbert space technique of Sec. \!IV.
In Sec. VIII we found all such affine quantum states that are finite in the usual $L^2$ norm.  However, the so-called
problem of time in quantum gravity leads one to consider alternative normalization choices, such as those given in
\cite{Kuchar,wald,Kij}.  In particular, as we have shown, the quantum Hamiltonian constraint equation leads to a hyperbolic
equation that is more like the wave equation than the usual Schr\"odinger equation of standard quantum mechanics.  Under
such circumstances, it is argued in \cite{wald} that it is more natural to use the Klein-Gordon norm rather than the
standard $L^2$ norm.  It is possible that for the processes relevant for the formation of spikes, the quantum Hamiltonian
constraint can be approximately solved in closed form.  But if not, then standard numerical methods used to treat hyperbolic
equations could be used instead.

\acknowledgments We would like to thank Abhay Ashtekar, Vladimir Belinski, Woei-Chet Lim, David Sloan, Claes Uggla, and Bob Wald for helpful discussions.  DG was supported by NSF grants PHY-1205202 and PHY-1505565 to Oakland University.

\section*{Appendix: Hermiticity of the affine Hamiltonian constraint}

Here, we give an outline of the proof that the operator $\hat{H}_C$, defined in Eq. \!\eqref{r2}, is symmetric on the space
of functions satisfying suitable boundary conditions or having compact support in $\dR^3$.

It is clear that the most problematic  terms in \eqref{r2} are the ones with the second partial derivatives, i.e., the second and the third
terms of the first line of \eqref{r2}.  One can easily show that when taken separately, each of them is not symmetric. In what follows
we show that the sum of them has however this property.  To demonstrate this, we make use of the following identity,
\begin{equation}\label{sym1}
  \sum_I x_I^2\frac{\partial^2}{\partial x_I^2} -  \sum_{I\neq J} x_I x_J \frac{\partial^2}{\partial x_I
\partial x_J} =  \sum_I H_I -  \sum_{I\neq J} H_{IJ}\, ,
\end{equation}
where
\begin{equation}\label{sym2}
H_I := x_I^2\frac{\partial^2}{\partial x_I^2} + 2 x_I \frac{\partial}{\partial x_I}
\end{equation}
and
\begin{equation}\label{sym3}
H_{IJ}:= x_I x_J \frac{\partial^2}{\partial x_I \partial x_J} + x_I \frac{\partial}{\partial x_I} \, .
\end{equation}
The proof consists in showing  that
\begin{equation}\label{sym4}
  \sum_I [\langle f| H_I  g \rangle - \langle H_I f|  g \rangle] = 0 = \sum_{I \neq J}\left[ \langle f| H_{IJ}
   g \rangle - \langle H_{IJ} f|  g \rangle\right].
\end{equation}
Making use of the  identities:

\begin{equation}\label{sym5}
  x^2_I\; \frac{\partial^2 f^\ast}{\partial x_I^2}\; g = \frac{\partial}{\partial x_I}\; (x^2_I\;\frac{\partial f^\ast}{\partial x_I}\; g)
  - 2 x_I\;\frac{\partial f^\ast}{\partial x_I}\;g - x^2_I\;\frac{\partial f^\ast}{\partial x_I}\;\frac{\partial g }{\partial x_I}\, ,
\end{equation}

\begin{equation}\label{sym6}
  x^2_I\;  f^\ast\; \frac{\partial^2 g}{\partial x_I^2} = \frac{\partial}{\partial x_I}\; (x^2_I\;f^\ast\;\frac{\partial g}{\partial x_I})
  - 2 x_I\;f^\ast\; \frac{\partial g}{\partial x_I} - x^2_I\;\frac{\partial f^\ast}{\partial x_I}\;\frac{\partial g }{\partial x_I}\, ,
\end{equation}
and

\begin{align}\label{sym7}
  x_I x_J\frac{\partial^2 f^\ast}{\partial x_J x_I}\; g &= \frac{\partial}{\partial x_J} (x_I x_J\frac{\partial f^\ast}{\partial x_I} g)
  -  x_I\frac{\partial f^\ast}{\partial x_I}g - x_I x_J\frac{\partial f^\ast}{\partial x_I}\frac{\partial g }{\partial x_J}\, ,\\
  x_I x_Jf^\ast\frac{\partial^2 g}{\partial x_I x_J} &= \frac{\partial}{\partial x_I} (x_I x_J f^\ast \frac{\partial g}{\partial x_I} )
  -  x_J f^\ast\frac{\partial g}{\partial x_J} - x_I x_J \frac{\partial f^\ast}{\partial x_I}\frac{\partial g }{\partial x_J}\, ,
\end{align}
one can find  that the integrants of  \eqref{sym4} consist entirely of the factors:
\begin{equation}\label{sym8}
 \frac{\partial}{\partial x_I} [x_I^2 (\frac{\partial f^\ast}{\partial x_I}\;g - f^\ast\;
\frac{\partial g}{\partial x_I})]~~~\mbox{and}~~~\frac{\partial}{\partial x_J} (x_I x_J\frac{\partial f^\ast}{\partial x_I} g)
-\frac{\partial}{\partial x_I} (x_I x_J f^\ast \frac{\partial g}{\partial x_I} )\, .
\end{equation}
Due to Eq.  \!\eqref{sym8} it is easy to show that Eq. \!\eqref{sym4} is satisfied in the subspace of functions with compact support
$C_0 (\dR^3)\subset L^2(\dR^3, d^3 x)$, or in the subspace of functions satisfying suitable boundary conditions.
In the rhs of \eqref{sym1} we have the cancellation of the linear terms of  \eqref{sym2} and \eqref{sym3}, which leads to
the lhs of \eqref{sym1}.

\end{document}